\newcommand{\apj}{ApJ}
\newcommand{\apjl}{ApJL}
\newcommand{\aap}{A\&A}

\documentclass{iaus}
\usepackage{graphicx}

\title[VLBA determinations of the distances to nearby star-forming regions]
{VLBA determinations of the distances to nearby star-forming regions}

\author[Loinard et al.]   
{Laurent Loinard$^1$, Rosa M.\ Torres$^1$,\\ Amy J.\ Mioduszewski$^2$ \and Luis F.\ Rodr\'{\i}guez$^1$}

\affiliation{$^1$Centro de Radiastronom\'{\i}a y Astrof\'{\i}sica,
Universidad Nacional Aut\'onoma de M\'exico \\
Apartado Postal 72-3 (Xangari), 58089 Morelia, Michoac\'an, M\'exico \\
email: {\tt l.loinard,r.torres,l.rodriguez@astrosmo.unam.mx} \\
[\affilskip]$^2$National Radio Astronomy Observatory,
Array Operations Center \\
1003 Lopezville Road, Socorro, NM 87801, USA \\
email: {\tt amiodusz@aoc.nrao.edu}}

\pubyear{2008}
\volume{248}  
\pagerange{119--126}
\jname{A Giant Step : from Milli- to Micro-arcsec Astrometry}
\editors{A.C. Editor, B.D. Editor \& C.E. Editor, eds.}
\begin{document}

\maketitle

\begin{abstract} Using phase-referenced multi-epoch {\em Very Long
Baseline Array} observations, we have measured the trigonometric
parallax of several young stars in the Taurus and Ophiuchus
star-forming regions with unprecedented accuracy. The mean distance to
the Taurus complex was found to be about 140 pc, and its depth around
20 pc, comparable to the linear extent of Taurus on the plane of the
sky. In Ophiuchus, 4 sources were observed so far. Two of them were
found to be at about 160 pc (the distance traditionally attributed to
Ophiuchus), while the other 2 are at about 120 pc. Since the entire
Ophiuchus complex is only a few parsecs across, this difference is
unlikely to reflect the depth of the region. Instead, we argue that
two physically unrelated sites of star-formation are located along the
line of sight toward Ophiuchus.

\keywords{astrometry, stars: distances, stars: formation, radio continuum: 
stars, stars: magnetic fields, radiation mechanisms: nonthermal, techniques: 
interferometric, binaries: general, ISM: clouds}

\end{abstract}

\firstsection 
\section{Introduction}

To provide accurate observational constraints for pre-main sequence
evolutionary models, and thereby improve our understanding of
star-formation, it is crucial to measure as accurately as possible the
properties (age, mass, luminosity, etc.) of individual young stars.
The determination of most of these parameters, however, depends
critically on the often poorly known distance to the object under
consideration. While the average distance to nearby low-mass
star-forming regions (e.g.\ Taurus or $\rho-$Ophiuchus) has been
estimated to about 20\% precision using indirect methods (Elias
1978a,b; Kenyon et al.\ 1994; Knude \& Hog 1998; Bertout \& Genova
2006), the line-of-sight depth of these regions is largely unknown,
and accurate distances to individual objects are still missing.  Even
the highly successful Hipparchos mission (Perryman et al.\ 1997) did
little to improve the situation (Bertout et al.\ 1999) because young
stars are still heavily embedded in their parental clouds and are,
therefore, faint in the optical bands observed by Hipparchos.

Low-mass young stars often generate non-thermal continuum emission
produced by the interaction of free electrons with the intense
magnetic fields that tend to exist near their surfaces (e.g.\
Feigelson \& Montmerle 1999). If the magnetic field intensity and the
electron energy are sufficient, the resulting compact radio emission
can be detected with Very Long Baseline Interferometers (VLBI --e.g.\
Andr\'e et al.\ 1992). The relatively recent possibility of accurately
calibrating the phase of VLBI observations of faint, compact radio
sources using nearby quasars makes it possible to measure the absolute
position of these objects (or, more precisely, the angular offset
between them and the calibrating quasar) to better than a tenth of a
milli-arcsecond (Brisken et al.\ 2000, 2002; Loinard et al.\ 2005,
2007; Torres et al.\ 2007; Xu et al.\ 2006; Hachisuka et al. 2006;
Hirota et al.\ 2007; Sandstrom et al.\ 2007). This level of precision
is sufficient to constrain the trigonometric parallax of sources
within a few hundred parsecs of the Sun (in particular of nearby young
stars) with a precision better than a few percents using multi-epoch
VLBI observations. With this goal in mind, we have initiated a large
project aimed at accurately measuring the trigonometric parallax of a
sample of magnetically active young stars in the most prominent and
often-studied northern star-forming regions within 1 kpc of the Sun
(Taurus, $\rho-$Ophiuchus, Perseus, Serpens, etc.) using the
10-element Very Long Baseline Array (VLBA) of the National Radio
Astronomy Observatory (NRAO). Here, we will summarize the results 
obtained so far in Taurus and Ophiuchus.

\section{Observations and parallax determination}

All the observations used here were obtained in the continuum at 3.6
cm (8.42 GHz) with the {\em Very Long Baseline Array} (VLBA) of the
{\em National Radio Astronomy Observatory} (NRAO). A total of 7
sources were studied so far: three in Taurus (T Tau, Hubble 4, and
HDE~283572), and four in Ophiuchus (S1, DoAr21, VSSG14, and WL5). In
all cases, between 6 and 12 observations spread over 1.5 to 2 years
were obtained. The data were edited and calibrated following the
standard VLBA procedures for phase-referenced observations (see
Loinard et al.\ 2007, and Torres et al.\ 2007 for details). 

The displacement of the sources on the celestial sphere is the
combination of their trigonometric parallax ($\pi$) and their proper
motions ($\mu$). For isolated sources (such as Hubble 4, and
HDE~283572 in our case), it is common to consider linear and uniform
proper motions, so the right ascension and the declination vary as a
function of time $t$ as:

\begin{eqnarray} 
\alpha(t) & = & \alpha_0+(\mu_\alpha \cos \delta) t + \pi f_\alpha(t) \label{uni1}\\%
\delta(t) & = & \delta_0+\mu_\delta t + \pi f_\delta(t), \label{uni2} 
\end{eqnarray}

\noindent where $\alpha_0$ and $\delta_0$ are the coordinates of the
source at a given reference epoch, $\mu_\alpha$ and $\mu_\delta$ are
the components of the proper motion, and $f_\alpha$ and $f_\delta$ are
the projections over $\alpha$ and $\delta$, respectively, of the
parallactic ellipse.

For sources in multiple systems, however, the proper motions are
affected by the gravitational influence of the other members of the
system. As a consequence, the motions are curved and accelerated,
rather than linear and uniform. If the orbital period is long compared
with the timespan covered by the observations (as will be the case for
T Tau --see Loinard et al.\ 2003), it is sufficient to include a
uniform acceleration in the fit. This leads to functions of the form:

\begin{table}
 \begin{center}
  \caption{Distances to the sources in Taurus and Ophiuchus}
  \label{tab1}
  \begin{tabular}{lccccccc}
\hline 
Complex & \multicolumn{3}{c}{Taurus} & \multicolumn{4}{c}{Ophiuchus} \\%
\hline
Source        & T Tau & Hubble 4 & HDE~283572 & S1 & DoAr21 & VSSG14 & WL5 \\%
Distance (pc) & 147.6 $\pm$ 0.6 & 132.8 $\pm$ 0.5 & 128.5 $\pm$ 0.6 & 116.9 $^{+7.2}_{-6.4}$ & 121.9 $^{+5.8}_{-5.3}$ & 165.6 $^{+6.2}_{-5.8}$ & 168.3 $^{+8.2}_{-9.3}$ \\%
\hline 
  \end{tabular}
 \end{center}
\end{table}

\begin{eqnarray}
\alpha(t) & = & \alpha_0+(\mu_{\alpha 0} \cos \delta) t + {1 \over 2} (a_\alpha\cos \delta) t^2 + \pi f_\alpha(t) \label{acc1}\\%
\delta(t) & = & \delta_0+\mu_{\delta 0} t + {1 \over 2} a_\delta t^2 + \pi f_\delta(t), \label{acc2}
\end{eqnarray}

\noindent where $\mu_{\alpha 0}$ and $\mu_{\delta 0}$ are the proper
motions at a reference epoch, and $a_\alpha$ and $a_\delta$ are the
projections of the uniform acceleration (see Fig.\ 1 for a comparison
between fits with and without acceleration terms). 

Finally, if a source is a member of a multiple system whose orbital
period is shorter than, or comparable with the timespan covered by the
observations (as will be the case for all Ophiuchus sources), then a
full Keplerian fit is needed, but additionnal observations are
required to properly constrain that fit. These observations are
currently being obtained for all Ophiuchus sources, but are not yet
analyzed. As a consequence, the fits presented bellow for the
Ophiuchus sources are based on Eqs.\ \ref{uni1} and \ref{uni2}.  The
resulting uncertainties will be much larger than for the sources in
Taurus because the orbital motions generate an unmodelled scatter
around the mean positions of the sources. These increased errors
should disappear once we include the new observations, and perform a 
full Keplerian fit.

\begin{figure}[t]
\begin{center}
 \includegraphics[width=3.0in,angle=270]{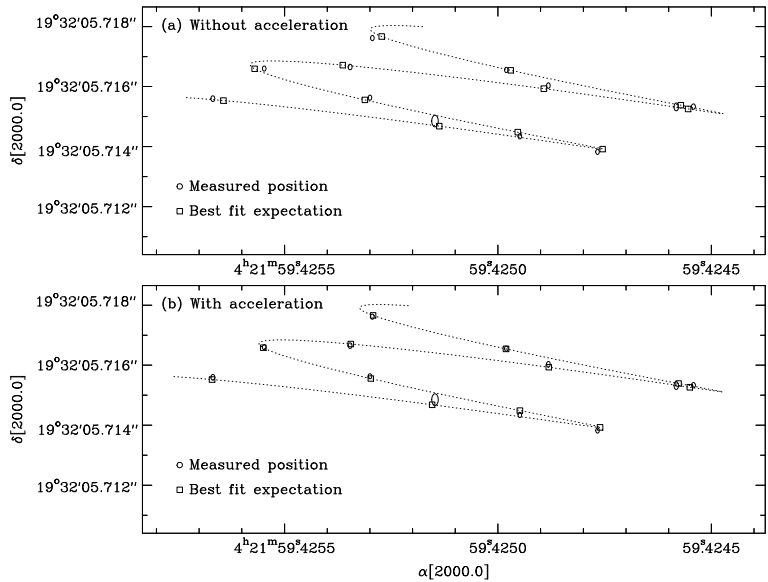} 
 \caption{Measured positions of T Tau Sb and best fit without (a) and
with (b) acceleration terms. The observed positions are shown as
ellipses, the size of which represents the magnitude of the
errors. Note the very significant improvement when acceleration terms
are included.}
   \label{fig1}
\end{center}
\end{figure}

\section{Results and Discussion}

The distance to the seven sources studied here are given in Tab.\ 1.
For the Taurus sources, Hipparchos parralaxes (Bertout et al.\ 1999),
and/or estimates based on a modified convergent point method (Bertout
\& Genova 2006) are available. Our measurements are always consistent
with these values, but are one to two orders of magnitude more
accurate. Only one other source in Taurus (V773 Tau; Lestrade et al.\
1999) has a VLBI-based parallax determination. Taking the mean of that
and our 3 measurements, we estimate the mean distance to the Taurus
cluster to be $\bar{d}$ = 137 pc. The dispersion about that mean leads
to a full width at half maximum depth of about 20 pc, comparable to
the linear extent of Taurus on the plane of the sky. 

Traditionnaly assumed to be at a distance of 165 pc (Chini 1981),
Ophiuchus has recently been proposed to be somewhat nearer, at 120 pc
(Knude \& Hog 1998). Interestingly, two of our sources are consistent
with the traditionnal value, but the other two are consistent with a
distance of 120 pc. Since Ophiuchus is only a few parsecs across on
the plane of the sky, it is very unlikely to be 40 pc deep. It is
noteworthy that the two sources at about 120 pc (S1 and DoAr21) are
associated with the sub-condensation Oph A, whereas the two sources at
$\sim$ 165 pc (VSSG14 and WL5) are associated with the condensation
Oph B (See Motte et al.\ 1998 for an overview of the Ophiuchus
complex). Thus, a plausible explanation of our results is that two
physically unrelated star-forming regions are located along the
line-of-sight toward Ophiuchus. Interestingly, Knude \& Hog (1998)
noticed an effect that would be consistent with this possibility in
their analysis of extinction towards Ophiuchus. While a clear
extinction step was clearly visible at 120 pc, extinction was
apparently extending up to about 160 pc. Additional VLBI observations
will be needed to confirm the existence of two regions of
star-formation towards Ophiuchus, and to investigate at which distance 
the other condensations in Ophiuchus are located.

\section{Conclusions and perspectives}

The present results show that VLBA observations of non-thermal sources
associated with young stars have the potential to improve very
significantly our knowledge of the space distribution of star-forming
regions in the Solar neighborod. Indeed, the precision obtained by
these measurements is even sufficient to probe the 3D structure of
star-forming regions. Coupled with pre-main sequence evolutionnary
models, such information could be used to reconstruct the history of
star-formation with individual regions.

\end{document}